\pdfoutput=1

\documentclass[table,11pt]{article}
\usepackage[table]{xcolor}
\usepackage[preprint]{acl}
\usepackage[table]{xcolor}

\usepackage{times}
\usepackage{latexsym}
\usepackage{comment}
\usepackage{graphicx}

\usepackage[T1]{fontenc}

\usepackage[utf8]{inputenc}

\usepackage{microtype}
\usepackage{pifont}
\usepackage{enumitem}

\usepackage{inconsolata}
\usepackage{array}

\usepackage[linesnumbered,ruled,vlined]{algorithm2e}
\usepackage{amsmath, amssymb}
\usepackage{float}
\usepackage{capt-of}

\SetCommentSty{mycommfont}

\SetKwInput{KwInput}{Input}    
\SetKwInput{KwOutput}{Output}  

\usepackage{tikz}       
\usepackage{pgfplots}   
\pgfplotsset{compat=1.17}
\usepackage{caption}    
\usepackage{amsmath}    
\usepackage{xcolor}     

\definecolor{mypurple}{RGB}{42,3,135}
\definecolor{myred}{RGB}{225,5,124}

\tikzset{
	RandRow/.style={mypurple,every mark/.append style={fill=mypurple},mark=otimes, mark size=2.5pt},
	RelRow/.style={myred,every mark/.append style={fill=myred},mark=oplus, mark size=2.5pt}
}

\usepackage{xspace}
\newcommand{\sys}{DBRAG\xspace}

\title{\sys: Multi-Table Retrieval-Augmented Generation for Complex Database Queries}

\author{Prince Larbi Ampofo \quad Ryoji Kubo \quad Djellel Difallah \\
  NYU Abu Dhabi, UAE \\
  \texttt{\{pla252,ryojikubo,djellel\}@nyu.edu}}

\usepackage{algorithmic}
\usepackage{amsmath}
\usepackage{tabularx}
\usepackage{adjustbox}
\usepackage{booktabs}
\usepackage{pgfplots}
\usepackage{subcaption}
\usepackage{multirow}
\usepackage{array}
\usepackage[normalem]{ulem}

\definecolor{ACLLinkBlue}{HTML}{000099}
\hypersetup{colorlinks=true, allcolors=ACLLinkBlue}

\newcommand{\ryoji}[1]{\textcolor{olive}{#1}}

\begin{document}
\maketitle

\begin{abstract}
Recent advancements in large language models have introduced new capabilities for reasoning over structured data, particularly through program-aided tools that can analyze tables. However, many existing methods address single-table scenarios or assume that the relevant tables are already provided. In practice, users often issue complex data exploration queries over entire databases, where relevant information may be distributed across multiple relations. In this work, we introduce DBRAG, a retrieval-augmented generation framework tailored for multi-table question answering. DBRAG first retrieves candidate tables using an offline table index, enriches their summaries with query-relevant rows, and uses an LLM to rerank the candidates. A program-aided reasoner then selects the required tables and executes operations over their full contents, keeping the initial prompt context compact. Experiments on the Spider, GeoQuery, and ATIS datasets used in this study demonstrate improvements in table retrieval and multi-table question answering.
\end{abstract}
\section{Introduction}
Tabular data structures are a widely used data model for storing information in various domains. In real-world applications, these tables often contain complex, interrelated data with relational dependencies. When interpreted, relational data can lead to improved decision-making, enhanced operational efficiency, and the discovery of new hidden insights~\cite{fey2023relationaldeeplearninggraph}. Given the inherent complexity of structured query languages, such as SQL, using natural language can lower the barrier for data exploration~\cite{DBLP:journals/pvldb/LiLCLT24}. Consequently, table question-answering (QA) has emerged as an important research area focused on developing systems that can understand natural language queries and extract relevant insights from tabular data~\cite{jin2022surveytablequestionanswering, Lu_2025}.

Although significant progress has been made in single-table QA scenarios~\citep{liu2022tapex, herzig-etal-2020-tapas, nan-etal-2022-fetaqa, zhu-etal-2021-tat}, the more challenging and practical task of multi-table QA remains relatively underexplored. Multi-table QA involves answering questions that require the integration of information from multiple tables, necessitating sophisticated reasoning and advanced data fusion techniques~\citep{pal-etal-2023-multitabqa}. This challenge is further compounded by the complex relationships between tables, such as identifying joinable keys or overlapping attributes, particularly when such relationships lack a well-structured explicit relational schema, as is often the case in web tables and data lakes~\cite{DBLP:journals/pvldb/CafarellaHWWZ08,DBLP:journals/vldb/ChapmanSKKIKG20}. 

Many existing studies assume that the relevant tables are already provided~\cite{chen2024tablerag, pal-etal-2023-multitabqa, wang2024chainoftable}. In an open-domain table QA setting, the system must instead retrieve them from a larger corpus. This requires considering both query-table relevance and table-table relationships: a table with limited direct relevance may still be needed to connect complementary information~\cite{chen2025tableretrievalsolvedproblem}.


Large language models (LLMs) have demonstrated remarkable proficiency in various natural language processing tasks~\cite{minaee2024largelanguagemodelssurvey}. Their adaptation to QA via retrieval-augmented generation (RAG) has led to substantial performance gains~\cite{DBLP:conf/nips/LewisPPPKGKLYR020}. In the RAG framework, relevant information is first efficiently retrieved from a pre-indexed large corpus and used to augment the query provided to the LLM, thereby mitigating issues such as hallucinations and outdated knowledge. 

Despite the effectiveness of the RAG framework in single-table QA contexts~\cite{chen2024tablerag}, its application to multi-table QA remains underexplored. The challenges of multi-table QA stem from several factors. First, multi-table QA requires the system to reason across multiple tables, often necessitating different types of joins and aggregations. Second, the sheer volume of tables in a typical database complicates the retrieval process, as the system must identify not only the main tables relevant to a query but also any auxiliary tables needed to connect and augment the data. Third, each table can vary significantly in size, both in the number of columns and rows. For instance, data warehouses often contain wide tables with billions of rows. These challenges, combined with the limitations of current LLM context windows, make the direct application of LLMs to multi-table QA non-trivial, thereby necessitating novel approaches.

To tackle the aforementioned challenges, we propose a novel methodology that extends the RAG framework to encompass the entire database. Our approach decomposes the problem of multi-table QA into two critical subtasks, \emph{multi-table retrieval} and \emph{multi-table reasoning}.
~\ding{192} In the retrieval step, we adopt a two-pronged approach to identify the relevant tables in a scalable manner. First, table search quickly selects a subset of tables relevant to the query from the entire database using a table index. Next, we enrich the tables of interest by augmenting them with query-relevant rows. The collected context is then fed into an LLM to select the most promising tables.
~\ding{193} In \emph{multi-table reasoning}, we propose a novel reasoning protocol designed to answer the natural language question given the retrieved tables. Specifically, we design a multi-table-aware Chain-of-Thought prompting technique combined with a program-aided reasoner that executes operations over the retrieved tables and returns the final answer.
Our experiments on multi-table QA datasets demonstrate that \sys not only improves the retrieval of relevant tables but also enhances the reasoning task, resulting in higher overall QA performance in the evaluated settings.

\noindent In summary, we make the following contributions:
\begin{itemize}[noitemsep,topsep=0pt,leftmargin=*,label=$\star$] 
\item We introduce a RAG approach for multi-table question answering in settings where the relevant tables must be retrieved from a large table corpus.
\item We propose \sys, a novel framework that decomposes multi-table QA into two subtasks: a two-stage retrieval strategy that leverages table indexing and context enrichment, and a multi-table reasoning protocol that integrates Chain-of-Thought prompting with program-aided execution. 
\item We demonstrate through comprehensive experiments that \sys improves both table retrieval and reasoning performance, resulting in effective end-to-end QA execution. 
\end{itemize}

\paragraph{Reproducibility}
Code, datasets, prompts, relevant parameters, and details of the experimental setup are available in the project repository:\\
\url{https://github.com/colab-nyuad/DBRAG}
\section{Related Work}

In this section, we provide a brief overview of retrieval-augmented generation and then discuss the most pertinent works on table QA, highlighting both the inspiration and the novelty of our contributions.

\paragraph{LLMs and RAG.} Large language models have revolutionized natural language processing with their advanced text comprehension and reasoning capabilities. However, they are not designed to handle domain-specific knowledge, rapidly changing information, private data, or long-tail entities. These limitations result in issues such as hallucinations~\cite{DBLP:journals/corr/abs-2311-05232}, spurious explanations~\cite{DBLP:phd/ethos/Camburu20}, and inconsistencies~\cite{DBLP:journals/tacl/ElazarKRRHSG21,DBLP:journals/corr/abs-2001-09977}. Recent techniques have improved reasoning through prompting methods, such as Chain-of-Thought (CoT)~\cite{DBLP:conf/nips/Wei0SBIXCLZ22} for step-by-step reasoning, and Self-Consistency~\cite{DBLP:conf/iclr/0002WSLCNCZ23} for selecting the most consistent answer across sampled reasoning paths.

To overcome these challenges, recent research has focused on injecting domain-specific knowledge into language models by integrating external sources and leveraging domain corpora. Retrieval-augmented generation (RAG) has proven effective thanks to a two-stage retrieve-and-read framework, in which the retriever quickly extracts relevant data from large corpora~\cite{DBLP:journals/corr/abs-2312-10997,DBLP:conf/kdd/FanDNWLYCL24}. By incorporating accurate, up-to-date information, RAG can enhance the LLM's response pertinence, reduce hallucinations, and produce more reliable, contextually relevant outputs.

\paragraph{Language Models for Table Understanding.}
Numerous studies have focused on pretraining table-specific language models (LMs) for downstream tasks. Building on the success of masked language modeling, TAPAS~\cite{herzig-etal-2020-tapas} introduces a BERT-like architecture enhanced with positional embeddings to encode table-specific information. During pretraining, the model reconstructs masked table elements. In contrast to TAPAS’s reconstruction approach, TAPEX~\cite{liu2022tapex} pretrains a  language model to execute SQL operations on tables, motivated by the intuition that accurate SQL execution reflects robust table comprehension. To advance LM capabilities in table retrieval, \citet{herzig-etal-2021-open} develop a dense retriever leveraging contrastive learning, pretraining, and negative sample mining. By employing TAPAS as the base encoder, they demonstrate that dense retrieval outperforms traditional methods like BM25~\citep{bm25} in open-domain single-table retrieval.

\paragraph{LLMs for Single-table QA.}
\citet{DATER} develop a method that enables an LLM to construct sub-evidences and sub-questions to answer natural language questions over large tables. Their approach employs in-context learning by providing a small set of manually crafted examples to guide table deconstruction and reasoning. \citet{chen2024tablerag} introduce a RAG framework for single-table QA, retrieving relevant schema and cells to address scalability and context length limitations when feeding large tables to the LLM. Inspired by the effectiveness of reasoning chains in textual contexts, \citet{wang2024chainoftable} propose using an LLM agent to dynamically plan table operations based on prior steps, forming an iterative reasoning process. They demonstrate that enabling LLMs to generate a reasoning chain improves performance in single-table QA.

\paragraph{Multi-table QA.}
While single-table QA has garnered research attention, studies on multi-table QA remain limited. \citet{pal-etal-2023-multitabqa} make a pioneering contribution by pretraining the TAPEX model on a multi-table dataset with SQL queries and tabular answers, followed by fine-tuning on natural language questions. However, their method assumes the relevant tables are already provided to the model and does not address scenarios where flattened table representations exceed the model’s maximum sequence length. \citet{chen2025tableretrievalsolvedproblem} introduce a method focused on multi-table retrieval, solving a mixed-integer program that jointly optimizes table-query relevance and table-table relevance. This approach adapts to databases lacking key-foreign-key constraints. It uses flattened table representations for coarse-grained retrieval, while downstream question answering relies on SQL generation and execution.

\section{Methodology}

\begin{figure*}[ht]
  \centering
  \includegraphics[width=\textwidth]{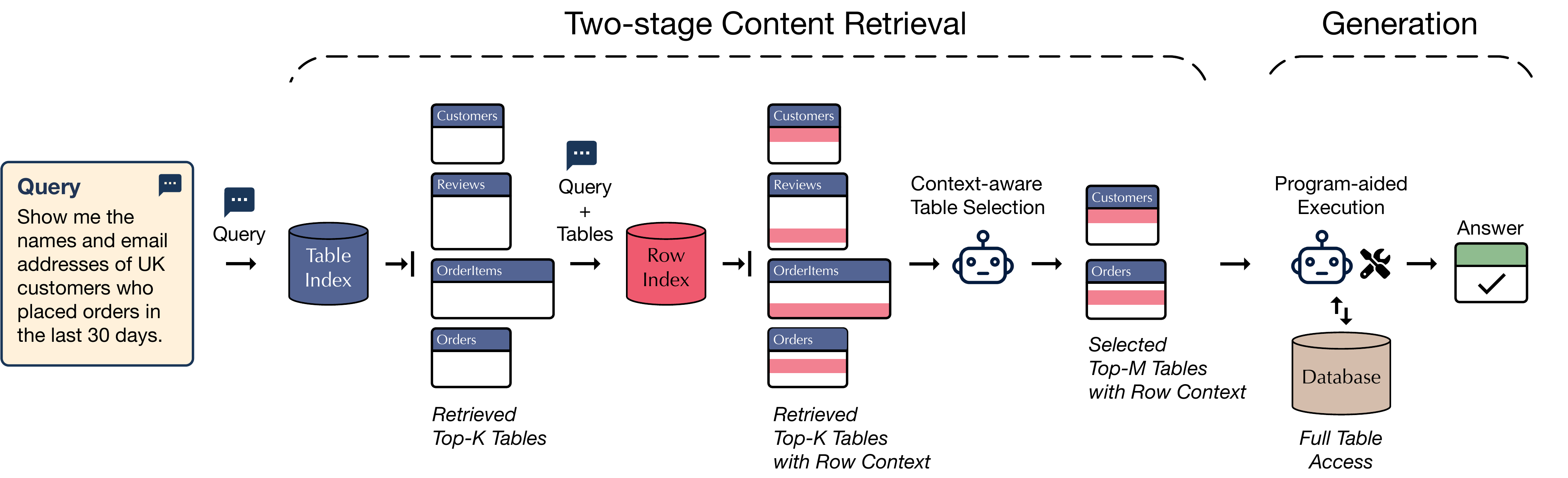}
  \caption{The \sys execution workflow. The Table Index retrieves top-$K$ candidates; the Row Index supplies query-relevant context rows. An LLM reranks the candidates to select top-$M$ tables. These tables are then passed to a program-aided reasoner, which uses a multi-table CoT prompt to generate the final answer.}
  \label{fig:dbrag}
\end{figure*}

\subsection{Motivation and Overview}
As mentioned in the introduction, directly presenting all tables to an LLM can be expensive and time-consuming. Additionally, methods that pass complete table information often run into context length limitations, especially with large tables. To make matters worse, users issuing ad hoc queries often do not know which tables are relevant or what the schema entails. To overcome these challenges and limit the number of tables presented to the LLM, we propose an approach that employs a multi-stage ranking mechanism for efficiency and mitigates context length issues by summarizing the selected tables. Additionally, we leverage a prompting technique that facilitates Chain-of-Thought-style multi-table reasoning, allowing the LLM to interact programmatically with the tables. Figure~\ref{fig:dbrag} shows an overview of our proposed framework, \sys. In the following, we begin by formally defining the setup, followed by descriptions of each component.

\subsection{Problem Formulation and Requirements} We formally define the multi-table QA task as follows. Given a natural language query \(q\) and a database \(\mathcal{D} = \{t_1, t_2, \ldots, t_N\}\), where each table \(t_i\) is described by a name \(t^{\text{Name}}_i\) and a schema \(t_i^{\text{Schema}} = \{c_1, c_2, \dots, c_n\}\), the objective is to produce an answer \(a\) that accurately addresses \(q\).

A standard RAG framework for multi-table QA can be decomposed into two stages:
~\ding{192} \emph{Multi-table QA Retrieval:} In this stage, the goal is to select a concise subset of tables, $\mathcal{R}_q \subset \mathcal{D}$, where $|\mathcal{R}_q| = M \ll N$. Ideally, this subset comprises only the tables necessary to answer the query.
~\ding{193} \emph{Multi-table QA Reasoner:} Here, an LLM is provided with the query ${q}$, the retrieved tables $\mathcal{R}_q$, and an appropriate prompt. The LLM processes this input to generate the final answer.

To develop an effective solution, we consider the following requirements: 
\begin{enumerate} [noitemsep,topsep=0pt,leftmargin=*,label=\textbf{(R\arabic*)}] 
\item \textbf{Retrieval Diversity:} Ranking tables independently by query similarity can overlook complementary tables needed for joins. The reranking prompt therefore asks the LLM to consider direct relevance, joinability, and collective coverage; this is a selection criterion rather than a formal diversity guarantee.
\item \textbf{Reasoner Context Size:} Full table contents can exceed the LLM context limit. Compact summaries guide table selection and operation construction, while execution tools access the full retrieved tables. Context-row selection therefore does not restrict answer computation to those rows.
\end{enumerate}

\subsection{Multi-table Retrieval}
\label{sec:retrieval}

To enable scalable retrieval of multiple relevant tables while addressing (\textbf{R1}), we decompose the retriever into a two-step process. First, an efficient encoder is used to perform coarse-grained ranking, identifying a broad subset of top-$K$ tables from the corpus. Then, an LLM is utilized to conduct fine-grained reranking of these top-$K$ tables to ensure higher precision in relevance and to identify the top-$M$ tables to form $\mathcal{R}_q$.

\paragraph{Table Index: Coarse-grained Ranking.}
We employ an efficient encoder \(f_{\text{enc}}\) to embed the coarse-grained summary of each table, which primarily consists of the table name, schema, and randomly sampled rows. For each table \(t_i\), we form its \emph{coarse-grained table summary} as follows:
\[
s_i = \{t^{\text{Name}}_i,\, t^{\text{Schema}}_i,\, t^{\text{RandomRows}}_i\},
\]
and compute its embedding \(\mathbf{e}_i = f_{\text{enc}}(s_i)\). The query \(q\) is embedded as \(\mathbf{q} = f_{\text{enc}}(q)\), and tables are ranked by the cosine similarity between \(\mathbf{q}\) and \(\mathbf{e}_i\). The top-\(K\) tables constitute the coarse-rank candidate set \(\mathcal{C}_q\). Since these embeddings are independent of the query, they can be precomputed and stored in a \emph{Table Index}.

\paragraph{Row Index: Context Retrieval.}
To further enrich the candidate set \(\mathcal{C}_q\), we augment each table with query-relevant row information. This process involves building a row embedding index by encoding each row in the database and storing it in a precomputed \emph{Row Index}. Specifically, for each row \(r\) with column names \(c_1, c_2, \dots, c_n\) and corresponding cell values \(v_1, v_2, \dots, v_n\), we form a row summary \(s^r = \{c_1, v_1, c_2, v_2, \dots, c_n, v_n\},\) and compute its embedding as \(\mathbf{e}^r = f_{\text{enc}}(s^r)\).

Adding the row context is crucial in DBRAG, as it enhances the system's understanding of which tables should be combined to answer the query. Accordingly, for each table \(t_i \in \mathcal{C}_q\), we query the Row Index to retrieve the rows most relevant to the query \(q\). Rows are ranked within each candidate table by the cosine similarity between $\mathbf{q}$ and $\mathbf{e}^r$. The selected rows replace the random rows in the summary, retaining the table name and schema, to form the \emph{fine-grained table summary}:
\[
s^+_i = \{t^{\text{Name}}_i,\, t^{\text{Schema}}_i,\, t^{\text{RelevantRows}}_i\},
\]
where \(t^{\text{RelevantRows}}_i\) denotes the set of $L$ relevant rows retrieved from the index.

\noindent\textbf{Note:} Although the Row Index can be queried directly to obtain a coarse-grained ranking, we use the Table Index to first narrow the candidate set. This design also allows alternative row context selections, such as \emph{Random Rows}, to be compared within the same retrieval framework (Section~\ref{sec:retrievalexp}).

\paragraph{LLM-based Fine-grained Reranking.}
Next, we employ a language model \(\mathcal{L}\) to perform fine-grained reranking of the candidate set \(\mathcal{C}_q\). This step is crucial for addressing (\textbf{R1}), as encoder-based coarse-grained ranking scores each table independently against the query. This focus on individual relevance can limit the selection of tables providing complementary information necessary for answering the query through joins and unions. The LLM-based reranker leverages its reasoning capabilities to consider relationships among the candidate tables, enabling the identification of relevant tables.

Given the query \(q\), a ranking prompt \(p_{\text{rank}}\), and the fine-grained table summaries \(\{s^+_i \mid t_i \in \mathcal{C}_q\}\), the LLM selects the top-\(M\) tables (with \(M < K\)) best suited to answer the query, denoted by \(\mathcal{R}_q\):
\[
\mathcal{R}_q = \mathcal{L}\Big(q,\, p_{\text{rank}},\, \{s^+_i \mid t_i \in \mathcal{C}_q\},\, M\Big).
\]
This reduced subset of tables, which is semantically relevant to the query and potentially complementary in providing the answer, is passed to the reasoning stage, which we explore next.

\subsection{Multi-table Reasoning}
\label{sec:reasoning}
To support queries that may require combining data from a varying number of tables while limiting context size (\textbf{R2}), we design a new multi-table reasoner. This component dynamically identifies the subset of tables needed from the retrieved set and generates an execution plan using an LLM integrated with execution tools to run code over the retrieved tables.

\paragraph{Table Selection.} First, the LLM is instructed to identify the specific subset of tables $\mathcal{R}^*_q$ from $\mathcal{R}_q$ that are necessary to answer the query. It dynamically evaluates the schema and context rows of each table using the set of table summaries:
\[
S^\mathcal{R}_q = \{ s^+_i \mid t_i \in \mathcal{R}_q \}.
\]
This step helps determine which tables can be effectively joined or unioned to construct the answer.

\paragraph{Program-aided CoT Reasoning.} Next, given the selected tables $\mathcal{R}^*_q$, the LLM generates a sequence of programmatic actions. Inspired by \citet{wang2024chainoftable}, we adopt a chain-of-thought prompting approach that decomposes the execution into a series of discrete data operations. To facilitate interaction with the tables, we integrate a programmatic function $\texttt{tool}$ into the prompt, which enables the LLM to perform iterative data manipulation on the tables.

At this stage, the LLM combines table selection with execution guidance in a unified prompt $p_{reason}$:
\[
ans = \mathcal{L}(q, p_{reason}, S^\mathcal{R}_q, \texttt{tool}).
\]
During the execution, intermediate results are
retained for use in subsequent operations, and once the chain
of operations reaches an end, the final answer is returned to the user.

\section{Experiments}

We evaluate our method and baselines focusing on two key tasks: multi-table retrieval and multi-table reasoning.

\paragraph{Datasets.}
Existing tabular QA benchmarks have largely focused on either limited-domain scenarios or single-table contexts. To address this gap, MTQA~\citep{pal-etal-2023-multitabqa} introduced a multi-table QA dataset derived from the cross-domain Spider dataset~\cite{yu-etal-2018-spider}, which spans tables from 138 distinct domains. Their work adapts the original text-to-SQL pairs by executing the SQL queries to obtain tabular answers and excluding examples with errors or empty results. In our experiments, we also use the MTQA versions of GeoQuery (GeoQ;~\citealp{10.5555/1864519.1864543}) and ATIS~\cite{price-1990-evaluation, dahl-etal-1994-expanding}, two smaller multi-table QA datasets built on databases of United States geography and flight booking systems, respectively. Appendix~\ref{ap:dataset} details the methodology for constructing the database corpora and the queries. It also includes a summary of the key statistics of each dataset.

\paragraph{Experimental Setup and Parameters.} 
\label{sec:default}
In our method, \sys, we utilize \texttt{GPT-3.5-turbo}\footnote{The specific version is \texttt{gpt-3.5-turbo-0125}. \url{https://platform.openai.com/docs/models}} as the backbone LLM. For efficient dense retrieval, we employ OpenAI's \texttt{text-embedding-3-small}\footnote{\url{https://platform.openai.com/docs/guides/embeddings}} as the encoder \(f_{enc}\). Specifically, we set the coarse-grained table budget to \(K = 50\) and the fine-grained table budget to \(M = 5\), a design choice that balances precision and recall across diverse user queries. Unless otherwise specified, the number of Row Context examples added to the table summaries is set to \(L = 2\). Additionally, all the LLM-based experiments are run with a self-consistency parameter of 5, where we take the majority vote from five independent runs.

\subsection{Multi-table Retrieval Performance}
\label{sec:retrievalexp}
We begin by evaluating the table retrieval performance. In the following experiments, we focus on the larger Spider dataset, which presents a more challenging retrieval setup, and omit GeoQ and ATIS; their limited number of tables consistently results in perfect recall across all baselines.

\begin{table*}[!ht]
\centering
\small
\resizebox{0.95\textwidth}{!}{
\begin{tabularx}{\textwidth}{l*{4}{>{\centering\arraybackslash}X}|>{\centering\arraybackslash}X}
\toprule
\multirow{2}{*}{Method} & \multicolumn{4}{c|}{Queries} & \multirow{2}{*}{Average} \\
                        & 1-table & 2-table & 3-table & 4-table &  \\
                        & (556)   & (366)   & (57)    & (6)     &  \\
\midrule
\multicolumn{6}{l}{\textit{Base Retrieval Methods}} \\
\midrule
Contriever         & 86.3   & 74.9   & 60.2  & 87.5   & 80.6  \\
DTR                & 90.5   & 75.6   & 57.9  & 83.3   & 83.0  \\
Coarse-rank        & 96.0 & 83.5 & 64.3 & 91.7  & 89.5 \\
\midrule
\multicolumn{6}{l}{\textit{Reranking Methods}} \\
\midrule
JAR (Contriever)   & 91.4   & 90.4   & 78.4  & \textbf{100}  & 90.3  \\
JAR (DTR)          & 91.2   & 83.5   & 70.2  & 91.7   & 87.1  \\
JAR (Coarse-rank)   & \underline{96.8} & 90.3   & 73.7  & 91.7   & 93.0  \\
\midrule
\sys (Name and Schema)     & 89.4   & 77.5   & 68.4  & 83.3   & 83.7  \\
\sys (Random Rows)         & 95.7   & \underline{94.7} & \textbf{87.7} & \textbf{100} & \underline{94.9} \\
\sys (Relevant Rows)       & \textbf{97.1} & \textbf{95.0} & \underline{87.1} & \textbf{100} & \textbf{95.8} \\
\bottomrule
\end{tabularx}
}
\caption{Multi-table retrieval performance (Recall@5 on Spider). Best result in \textbf{bold}, second-best result \underline{underlined}. Query counts (by relevant tables) in parentheses.}
\label{table:retrieval_result}
\end{table*}

\paragraph{Baselines.}

Following the work of~\citet{chen2025tableretrievalsolvedproblem} on table retrieval, we consider the following baseline methods:

\begin{itemize}[noitemsep,topsep=0pt,leftmargin=*]
\item \text{Contriever}~\cite{izacard2022unsupervised}: A text retrieval method that does not incorporate any table-specific model design. It ranks tables based on the cosine similarity between the embeddings of the question and the flattened tables. 
\item \text{DTR}~\cite{herzig-etal-2021-open}: A method specifically designed for table retrieval. TAPAS-large is fine-tuned under this framework, and the relevance of each table to the question is computed accordingly. 
\item \text{Coarse-rank:} We use our encoder-based coarse ranking (see Section~\ref{sec:retrieval}) and select the top \(K=M=5\) tables, thereby bypassing the LLM-based fine-grained reranking step.
\item \text{JAR}~\cite{chen2025tableretrievalsolvedproblem}: This method employs a base retrieval method and solves a mixed-integer program to identify relevant tables. We evaluate three versions: (i) JAR (Contriever), (ii) JAR (DTR), and (iii) JAR (Coarse-rank), which use the JAR algorithm to rerank the results of Contriever, DTR, and our coarse-rank approach, respectively. 
\end{itemize}

\paragraph{\sys Ablations.} We evaluate several variations of our method to analyze the contribution of each component. 
\begin{itemize}[noitemsep,topsep=0pt,leftmargin=*]
\item \sys(Name and Schema): Modifies the fine-grained summary so that it includes only the table name and the table schema.
\item \sys(Random Rows): Modifies the fine-grained summary to include the table name, table schema, and random rows from the table. In this setup, the LLM-based reranking receives random rows instead of the relevant rows.
\item \sys(Relevant Rows): This is our default setup described in Section~\ref{sec:default}.
\end{itemize}

\paragraph{Evaluation Metrics.}
To assess the quality of multi-table retrieval, we compare the retrieved tables against the relevant ground truth tables. Specifically, we use Recall@5 as the metric to evaluate retrieval performance. Recall@5 measures the proportion of required tables retrieved among the top five candidates, averaged across queries. For a fair evaluation, we retrieve the top $M = 5$ tables from each method as the set of retrieved tables.

\paragraph{Retrieval Results \& Discussion.}
The results of multi-table retrieval are presented in Table~\ref{table:retrieval_result}, with detailed performance metrics for queries involving different numbers of relevant tables. It can be observed that as the number of relevant tables required to answer a question increases, retrieval performance generally decreases. However, queries with four relevant tables show improved performance, although this group contains only six queries. Our method, \sys, which uses fine-grained table summaries with relevant rows, achieves the best overall performance.

The comparison with Coarse-rank highlights the significance of LLM-based reranking in improving retriever performance. Additionally, the comparisons with \sys~(Name and Schema) demonstrate the importance of incorporating row information alongside the table name and schema for accurate retrieval. Finally, the comparison with \sys~(Random Rows) underscores the effectiveness of retrieving relevant rows in enhancing retrieval performance.


\begin{table}[!ht]
\centering
\small
\resizebox{\columnwidth}{!}{
\begin{tabular}{lccc}
\toprule
Query & Llama 3.3 & Qwen2.5 & GPT-3.5 \\
      & {\scriptsize (70B-Instruct)} & {\scriptsize (32B-Instruct)} & {\scriptsize (Turbo-0125)} \\
\midrule
1-table (556)  & \underline{96.9}   & 96.4   & \textbf{97.1} \\
2-table (366)  & \underline{88.7} & 84.6   & \textbf{95.0}   \\
3-table (57)   & 73.1 & \underline{76.6}   & \textbf{87.1}   \\
4-table (6)    & \underline{91.6} & 83.3   & \textbf{100}    \\
\midrule
Average        & \underline{92.5} & 90.8   & \textbf{95.8}   \\
\bottomrule
\end{tabular}
}
\caption{Fine-grained reranking with different LLMs (Recall@5 on Spider; query counts in parentheses).}
\label{table:retrieval_result_LLM}
\end{table}

\paragraph{Fine-grained Reranking with Different LLMs.} In this experiment, we evaluate the performance of \sys in fine-grained reranking using three different LLMs, while fixing the default parameters (see Section~\ref{sec:default}). We report Recall@5 on the Spider dataset. Table~\ref{table:retrieval_result_LLM} shows that GPT-3.5 delivers the best results, especially as query complexity increases (i.e., when more tables are required). Although Llama 3.3~\citep{llama} and Qwen2.5~\citep{qwen} perform competitively on single-table queries, they do not keep up with multi-table scenarios. These results underscore the importance of advanced reasoning for effective fine-grained reranking for complex queries. These results show that the choice of reranking model affects performance under the fixed retrieval setup.

\subsection{Multi-table Reasoning Performance}
\label{sec:reasoningexp}
Next, we turn our attention to evaluating the end-to-end performance of multi-table QA encompassing retrieval and reasoning capabilities. In the following, we use the set of retrieved tables $\mathcal{R}_q$ obtained using \sys during the multi-table retrieval phase for the three datasets, and feed them to several baselines.

\paragraph{Baselines.} We compare our method against the following baselines:
\begin{itemize}[noitemsep,topsep=0pt,leftmargin=*]
    \item \text{ReadTable}~\cite{chen2024tablerag}: This is a common baseline that embeds the entire table in the prompt and asks the LLM to solve the problem using that data. However, if the table exceeds the model's context length, the question is marked as a failure.
    \item \text{MTQA}~\cite{pal-etal-2023-multitabqa}: A multi-table reasoning baseline that pretrains the TAPEX model on a multi-table dataset containing SQL queries and tabular answers, and then fine-tunes it on natural language questions.
\end{itemize}

\paragraph{Ablations.} We compare our method with the following variations:
\begin{itemize}[noitemsep,topsep=0pt,leftmargin=*]
    \item \sys (Name and Schema): This setup restricts the table summaries to include only the table name and schema, without context rows.
    \item \text{\sys (w/o Selection)}: Our approach that performs program-aided reasoning without the table-selection step (Section~\ref{sec:reasoning}).
    \item \text{\sys (Random Rows)}: Here, the relevant rows in the table summaries are replaced with random rows.
\end{itemize}


\begin{table*}[!ht]
\centering
\begin{small}
    \begin{adjustbox}{max width=\textwidth}
    \begin{tabular}{ll|c|ccc|ccc|ccc}
    \toprule 
    \multirow{2}{*}{Dataset} & \multirow{2}{*}{Method} & \multirow{2}{*}{Table EM}  & \multicolumn{3}{c|}{Row EM}    & \multicolumn{3}{c|}{Column EM} & \multicolumn{3}{c}{Cell EM} \\
    \cmidrule(lr){4-6} \cmidrule(lr){7-9} \cmidrule(lr){10-12}
     &  &  &  P & R & F1                   & P & R & F1                    &  P & R & F1 \\
    \midrule
    \multirow{6}{*}{Spider} 
     & ReadTable                  & 29.6 & 35.6  & 35.9 & 35.8 & 33.6  & 33.8 & 33.7 & 40.1 & 41.3 & 40.7  \\
     & MTQA                       & 10.6 & 22.0 & 22.6 & 22.3 & 12.6 & 12.6 & 12.6 & 28.4 & 28.2 & 28.3 \\
     & \sys (w/o Selection)       & 43.2 & 50.3 & 49.5 & 49.9 & 49.7 & 50.4 & 50.1 & 59.7 & 59.4 & 59.6 \\
     & \sys (Name and Schema)     & 41.3 & 48.5 & 47.5 & 48.0 & 47.5 & 48.1 & 47.8 & 57.9 & 56.2 & 57.0 \\
     & \sys (Random Rows)        & \underline{44.4} & \underline{51.7} & \underline{50.8} & \underline{51.3} & \underline{50.3} & \underline{50.7} & \underline{50.5} & \underline{60.9} & \underline{59.5} & \underline{60.2} \\
     & \sys (Relevant Rows)      & \textbf{44.9} & \textbf{52.7} & \textbf{51.9} & \textbf{52.3} & \textbf{51.3} & \textbf{51.8} & \textbf{51.5} & \textbf{62.6} & \textbf{61.6} & \textbf{62.1} \\
    \midrule
    \multirow{6}{*}{GeoQ} 
     & ReadTable                  & 42.3 & 55.3  & 50.1  & 52.6 & 50.1 & 58.9 & 54.1  & 63.9 & 68.2 & 66.0 \\
     & MTQA                       & \underline{47.2} & 55.8 & \underline{53.9} & 54.8 & 47.2 & 47.2 & 47.2 & 55.8 & 53.9 & 54.8 \\
     & \sys (w/o Selection)       & 43.5 & 57.5 & 53.5 & \underline{55.4} & 50.2 & 58.9 & 54.2 & 66.0 & \underline{70.5} & \underline{68.2} \\
     & \sys (Name and Schema)     & 31.6 & 44.8 & 39.6 & 42.0 & 36.4 & 42.3 & 39.1 & 50.3 & 51.0 & 50.6 \\
     & \sys (Random Rows)        & 45.1 & \underline{58.5} & 52.6 & \underline{55.4} & \underline{51.8} & \underline{59.3} & \underline{55.3} & \underline{66.3} & 68.2 & 67.2 \\
     & \sys (Relevant Rows)      & \textbf{47.8} & \textbf{60.3} & \textbf{55.0} & \textbf{57.5} & \textbf{54.6} & \textbf{62.9} & \textbf{58.4} & \textbf{68.7} & \textbf{71.2} & \textbf{69.9} \\
    \midrule
    \multirow{6}{*}{ATIS} 
     & ReadTable                  & \underline{80.2} & 85.2  & 86.1  & 85.6 & \underline{81.9} & \underline{84.9} &\underline{83.4} & 87.0  & \underline{91.9} & \underline{89.4} \\
     & MTQA                       & 75.6 & 82.4 & \underline{90.4} & 86.2 & 75.7 & 76.7 & 76.2 & 82.5 & 91.6 & 86.8 \\
     & \sys (w/o Selection)       & 76.7 & 83.6 & 84.0 & 83.8 & 77.4 & 79.1 & 78.3 & 85.0 & 89.8 & 87.3 \\
     & \sys (Name and Schema)     & 65.1 & 71.6 & 73.3 & 72.4 & 66.9 & 68.6 & 67.7 & 73.5 & 77.9 & 75.7 \\
     & \sys (Random Rows)        & \underline{80.2} & \underline{87.7} & 87.5 & \underline{87.6} & 81.0 & 82.6 & 81.8 & \underline{88.7} & 90.1 & \underline{89.4} \\
     & \sys (Relevant Rows)      & \textbf{84.9} & \textbf{91.2} & \textbf{91.9} & \textbf{91.5} & \textbf{85.3} & \textbf{86.1} & \textbf{85.7} & \textbf{92.3} & \textbf{93.9} & \textbf{93.1} \\
    \bottomrule
    \end{tabular}
    \end{adjustbox}
\end{small}
\caption{Performance evaluation of multi-table reasoning. Best results in \textbf{bold} and second-best results \underline{underlined}.}
\label{table:reasoning_result}
\end{table*}

\paragraph{Evaluation Metrics.}
The performance of multi-table reasoning is measured by comparing the ground truth answer table derived from the answer SQL query with the table generated by the model. Following \citet{pal-etal-2023-multitabqa}, we employ metrics that determine if the returned table's features precisely match those of the ground truth answer.

The evaluation metrics assess table similarity at different levels of granularity. \textbf{Table EM} measures whether the returned table exactly matches the target table, requiring both row and column ordering to be identical; since many queries do not depend on ordering, additional metrics are used. \textbf{Row EM} assesses how many rows in the returned table exactly match those in the target table, ignoring row order but requiring exact value ordering within each row. Similarly, \textbf{Column EM} evaluates how many columns in the returned table match those in the target table, disregarding column order while still requiring exact value ordering within each column. Finally, \textbf{Cell EM} provides the most relaxed evaluation by measuring how many individual cells match exactly between the two tables.

For Row EM, Column EM, and Cell EM, we report the macro versions of these metrics to mitigate the bias introduced by queries returning large tables. Statistics about table size and distribution are provided in Appendix~\ref{ap:dataset}.

\paragraph{Reasoning Results.}
Table~\ref{table:reasoning_result} presents the multi-table reasoning results on Spider, GeoQ, and ATIS. We observe that \sys outperforms MTQA and ReadTable on most metrics, particularly on Spider, where many queries fail because their tables exceed the LLM's context limit when using ReadTable. In our evaluations, MTQA shows lower performance than \sys across all datasets, which may stem from its lack of an explicit table selection step when handling complex multi-table queries. \sys's improvements on GeoQ and ATIS further validate its efficacy on domain-specific tasks, although simpler methods like ReadTable perform reasonably well in these setups.

Finally, the ablation studies confirm the importance of the main components of \sys. For instance, \sys outperforms its \sys~(w/o Selection) alternative, and comparisons with the \sys~(Name and Schema) and \sys~(Random Rows) versions show that including relevant rows in table summaries improves answer accuracy.

\subsection{Varying Table and Row Context Budget}
\begin{figure}[!ht]
    \centering
    \begin{minipage}[b]{0.48\linewidth}
        \centering
        \begin{tikzpicture}
            \begin{axis}[
                ybar,
                ylabel near ticks,
                width=1.11\linewidth,
                height=0.9\linewidth,
                bar width=5pt,
                xlabel={(a) Retrieved Tables ($K$)},
                ylabel={Recall@5},
                ymin=85, ymax=97,
                ytick={85, 90, 95, 100},
                ylabel style={yshift=-1ex,font=\small},
                xlabel style={yshift=0ex, font=\small},
                symbolic x coords={5, 10, 20, 50},
                xtick=data,
                ymajorgrids=true,
                grid style=dashed,
                legend style={at={(1.2,1.4)}, anchor=south, font=\LARGE, legend columns=2},
                yticklabel style={font=\small},
                xticklabel style={font=\small},
                legend style={at={(1.1, 1.15)}, legend columns=2, nodes={scale=0.5, transform shape}, style={column sep=0.25cm}},
            ]
            \addplot coordinates {(5, 89.5) (10, 94.2) (20, 94.5) (50, 95.8)};
            \addplot coordinates {(5, 89.5) (10, 93.3) (20, 95.1) (50, 94.9)};
            \addlegendentry{Relevant Rows}
            \addlegendentry{Random Rows}
            \end{axis}
        \end{tikzpicture}
    \end{minipage}\hspace{0.005\linewidth}
    \begin{minipage}[b]{0.48\linewidth}
        \centering
        \begin{tikzpicture}
            \begin{axis}[
                ybar,
                ylabel near ticks,
                width=1.1\linewidth,
                height=0.9\linewidth,
                bar width=5pt,
                xlabel={(b) Row Context Size ($L$)},
                ylabel={Table EM (\%)},
                ymin=40, ymax=48,
                ytick={42, 45, 48},
                ylabel style={yshift=-1ex,font=\small},
                xlabel style={yshift=0ex, font=\small},
                ymajorgrids=true,
                grid style=dashed,
                symbolic x coords={1, 2, 5, 10, 20},
                xtick=data,
                yticklabel style={font=\small},
                xticklabel style={font=\small},
                legend style={at={(1.1, 1.15)}, legend columns=2, nodes={scale=0.5, transform shape}, style={column sep=0.25cm}},
            ]
            \addplot coordinates {
          (2, 44.9) 
          (5, 46.0) 
          (10, 46.7)
          (20, 46.2)
        };
                    \addplot coordinates {
          (2, 44.4) 
          (5, 46.3) 
          (10, 46.3)
          (20, 45.1)
        };
            \end{axis}
        \end{tikzpicture}
    \end{minipage}
    \caption{\sys parameter variation on (a) retrieval and (b) reasoning on the Spider dataset.} 
    \label{fig:combined}
    \vspace{-0.2cm}
\end{figure}
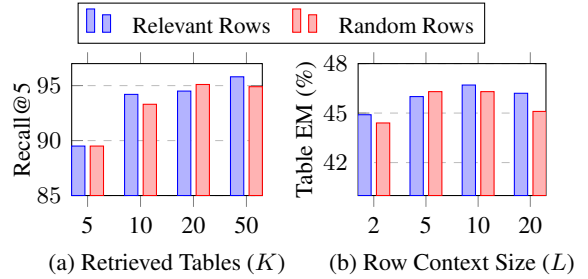

Finally, we evaluate the impact of table and row context budgets on both table retrieval and reasoning performance on the Spider dataset. In the retrieval setup (Fig.~\ref{fig:combined}(a)), we vary the coarse-grained ranking budget \(K\), which determines the number of retrieved tables provided to the LLM for fine-grained reranking. Our results indicate that increasing \(K\) generally improves retrieval performance for both approaches, and importantly, the setup using relevant row context mostly outperforms that using random row context. For the reasoning setup (Fig.~\ref{fig:combined}(b)), we vary the number of rows \(L\) included in the table summary. The experiment reveals that while increasing \(L\) initially enhances answer accuracy, performance reaches a plateau beyond which additional rows do not consistently improve performance. This aligns with the intuition that only a few rows are needed to grasp the structure and semantics of tabular information. These findings underscore the effectiveness of compact table summaries in the evaluated setting.

\section{Conclusions}
We propose DBRAG, a retrieval-augmented generation framework tailored for multi-table question answering. DBRAG overcomes the challenges of complex multi-table relationships by decomposing the task into efficient table retrieval and multi-table reasoning. Our approach leverages a two-stage retrieval strategy, first retrieving candidate tables from the Table Index and then reranking them using query-relevant row context, followed by multi-table chain-of-thought prompting and program-aided execution for precise reasoning. Extensive experiments on multi-table QA datasets demonstrate improvements in both table retrieval and end-to-end QA performance. Our results validate DBRAG as an effective solution for complex, multi-table database question answering. Future work will explore further optimizations and extend support for heterogeneous data integration such as semi-structured data and knowledge graphs.
\section{Limitations}
Building and maintaining the Row Index can be computationally expensive for databases with billions of records. More efficient index construction and maintenance strategies, including sampling approaches potentially drawing on the frequency-based cell selection used by \citet{chen2024tablerag}, could help address these challenges. We leave these issues, along with dynamic schema updates and production deployment, for future work.

While the queries in the datasets involve multiple tables, their answers can be obtained by executing SQL queries. In real-world scenarios, however, users may pose more complex queries, such as queries for open-ended exploration or interpretation of trends in the data, which may require analytical reasoning beyond SQL-based operations~\cite{ma-etal-2023-insightpilot}. The capability of DBRAG to handle such advanced queries remains unexplored in this paper.

Additionally, real-world large databases often present challenges such as noisy data, inconsistencies, and missing information~\cite{10.1109/TKDE.2023.3270101}. Our method has not been evaluated in these settings, where the model would need to navigate and address potential flaws in the data itself.

Finally, our comparisons cover a limited set of baselines for multi-table retrieval and reasoning. We do not include comparisons with advanced solvers which could provide further insights into the performance of our approach.

\bibliography{custom}
\appendix
\section{Datasets}
\label{ap:dataset}

\citet{pal-etal-2023-multitabqa} introduced a multi-table QA dataset based on the cross-domain Spider dataset, which comprises tables from 138 domains \cite{yu-etal-2018-spider}. Their work adapts the original text-to-SQL pairs by executing the SQL queries to obtain tabular answers and excluding examples with errors or empty results. Each sample includes a natural language question paired with its corresponding SQL query answer. Furthermore, we utilize GeoQuery (GeoQ;~\citealp{10.5555/1864519.1864543}) and ATIS~\cite{price-1990-evaluation, dahl-etal-1994-expanding}, which are multi-table QA datasets built on a database of United States geography and a relational database for flight booking systems, respectively.
In this work, we adopt \citet{pal-etal-2023-multitabqa}'s version
of these datasets, which uses \citet{iyer-etal-2017-learning}'s
text-to-SQL samples for GeoQ and ATIS.
In this version, the ATIS and GeoQ datasets include many unintelligible questions provided as disconnected sentences (e.g., “list la”). Consequently, we synthesized natural queries for the ATIS and GeoQ datasets based on the corresponding SQL queries. These queries, along with the code used for their generation, are available in the project repository.
Table~\ref{datasets} summarizes the statistics for each dataset.

\begin{table}[!ht]
    \centering
    
    \resizebox{0.9\columnwidth}{!}{%
    \begin{tabular}{lccc}
        \toprule
                           & \textbf{ATIS} & \textbf{GeoQ} & \textbf{Spider} \\
        \midrule
        \# train+valid samples   & 429         & 579         & 6715          \\
        \# test samples          & 86          & 253         & 985           \\
        \# tables                & 11          & 7           & 729           \\
        \bottomrule
    \end{tabular}%
    }
    \caption{Dataset Statistics}
    \label{datasets}
\end{table}

Table~\ref{tab:single_column_stats} presents an overview of additional dataset statistics of the three datasets focusing on their content. For each dataset, we report the total and unique number of questions, details on table joins (ranging from single-table to five-table questions), and summary statistics on table dimensions including the average, minimum, and maximum number of rows and columns. 

\begin{table}[!ht]
    \centering
    \resizebox{\columnwidth}{!}{%
    \begin{tabular}{lccc}
        \toprule
        \textbf{Metric} & \textbf{ATIS} & \textbf{GeoQ} & \textbf{Spider} \\
        \midrule
        Total questions & 515    & 832    & 7700 \\
        Total unique questions  & 422    & 737    & 7656 \\
        Total unique tables (name)    & 11     & 7      & 571 \\
        Total unique tables (content) & 11     & 7      & 729 \\
        \midrule
        1-table questions       & 190    & 717    & 4387 \\
        2-table questions       & 180    & 110    & 2592 \\
        3-table questions       & 79     & 5      & 655 \\
        4-table questions       & 66     & 0      & 56 \\
        5-table questions       & 0      & 0      & 10 \\
        \midrule
        Avg. no. of rows        & 82.29  & 175.36 & 1165.35 \\
        Min. no. of rows        & 14     & 32     & 2 \\
        Max. no. of rows        & 374    & 386    & 510437 \\
        \midrule
        Avg. no. of columns     & 5.3    & 4.4    & 5.99 \\
        Min. no. of columns     & 2      & 2      & 2 \\
        Max. no. of columns     & 15     & 6      & 48 \\
        \bottomrule
    \end{tabular}%
    }
        \caption{Additional Dataset Statistics}
    \label{tab:single_column_stats}

\end{table}

\section{Evaluation Metrics Details}
\label{ap:metrics}
This section complements the reasoning metrics described in Section~\ref{sec:reasoningexp}. For Row EM, Column EM, and Cell EM, \textit{precision} is the fraction of features (rows, columns, or cells) in the returned table that exactly match a feature in the target table, and \textit{recall} is the fraction of features in the target table that are exactly matched in the returned table; the F1-score is their harmonic mean. \citet{pal-etal-2023-multitabqa} compute these scores by pooling features across all returned and target tables in the test set (i.e., micro-averaging), so a few queries with very large returned or target tables can dominate the results. We instead compute precision and recall for each query, average them across queries (i.e., macro-averaging) so that every query contributes equally, and report F1 as the harmonic mean of the macro-averaged precision and recall.

\onecolumn
\section{\sys Prompts}
\label{sec:prompts}

\begin{table*}[!ht]
  \centering
  \resizebox{0.88\textwidth}{!}{%
  \renewcommand{\arraystretch}{1.25} 
  \ttfamily
  \begin{tabular}{|p{\textwidth}|}
  \hline
  \rowcolor{lightgray}
  \textbf{Reranking Prompt} \\ \hline
  \begin{minipage}[t]{\textwidth}
  \small

  Given a question and a set of candidate tables, rank tables by relevance to the given question while also considering their potential to be joined with others to enhance the completeness of the answer. Each table has a table\_id, table\_name, columns, and some rows (optional).

\bigskip
  - Question -\\[0.5em]
  \{question\}

\bigskip
  - Candidate Tables -\\
  \{candidate\_tables\}

\bigskip
  - Considerations for Ranking -
  \\
  1. Direct Relevance: Does the table contain column names or row values (if available) that strongly align with the terms/entities in the question?\\[0.5em]
  2. Joinability: Can the table be meaningfully joined with others based on shared column names or overlapping row values (if available)?\\[0.5em]
  3. Coverage \& Completeness: Does the table alone or when joined with others provide a comprehensive answer to the question?

  \bigskip
  - Requirements -\\
  1. Always return exactly \{M\} unique table\_ids\\[0.5em]
  2. Order from most to least relevant\\[0.5em]
  3. Consider both standalone value and join potential\\[0.5em]
  4. Prioritize direct relevance over theoretical join possibilities
  
\bigskip
  - Response Format -\\
  \texttt{[table\_id1, table\_id2, table\_id3, table\_id4, table\_id5]}\\
  \end{minipage} \\ \hline
  \end{tabular}%
}

  \caption{Fine-grained reranking prompt, including relevance, joinability, and coverage criteria.}
  \label{tab:retrieval} 

  \vspace{2em} 

  \resizebox{0.88\textwidth}{!}{%
  \renewcommand{\arraystretch}{1.25} 
  \ttfamily
  \begin{tabular}{|p{\textwidth}|}
    \hline
    \rowcolor{lightgray}
    \textbf{Reason Prompt} \\ \hline
    \begin{minipage}[t]{\textwidth}
      \small
    You are working with \{retrieved\_dfs\} dataframes in Python named df1, df2, and so on. Think step by step, utilizing the tools provided to answer the given question by executing a series of dataframe manipulation operations. Always start by identifying the appropriate dataframes. Then, create a chain of actions and execute it on the appropriate dataframes with the execute\_dataframe\_code tool. Finally, use the final\_structured\_output tool in your last call to return the final dataframe with sql-style column names.
      
      \bigskip
      
      Example chain input format:\\
      <BEGIN> -> action1 -> ... action2 -> action3 -> <END>
      
      \bigskip
      
      Always continue the chain with the above format for example:\\
      df\_dic['df1'].merge(df\_dic['df2'], on='personId') -> inter.mean(axis=1) -> <END>
      
      \bigskip
      
      Always refer to your dataframes as df\_dic[dataframe\_name].\\
      For example, instead of writing \texttt{df1.groupby(...)} you should write \texttt{df\_dic['df1'].groupby(...)}.\\
      If you continue from the current state of the dataframe, refer to it as \texttt{inter}.
      
      \bigskip
      
      - Example 1 -\\
      Question: List the creation year, name and budget of each department.\\
      Chain Logic: We first need to select the appropriate dataframe(s), join dataframe(s) if necessary, then select the creation year, name and budget columns.
      
      \bigskip
      
      - Example 2 -\\
      Question : What are the names of the states where at least 3 heads were born?\\
      Chain Logic: We first need to select the appropriate dataframe(s), join dataframe(s) if necessary, then group by the state names, count the number of heads born in each state and filter for states with at least 3 heads born.
      
      \bigskip
      
      Use the following tools: execute\_dataframe\_code, final\_structured\_output
      
      \bigskip
      
      \{finegrained\_table\_summaries\}
      
      \bigskip
      
      Always think step by step. Begin!
      \\
    \end{minipage} \\ \hline
  \end{tabular}%
}

  \caption{Program-aided multi-table reasoning prompt using Python dataframe tools.}
  \label{tab:reason} 
\end{table*}

\end{document}